\documentclass[12pt]{iopart}

\usepackage[english]{babel}
\usepackage{graphicx}
\usepackage{units}
\usepackage{amsbsy,amssymb}
\usepackage{setstack}

\usepackage{color}

\usepackage[normalem]{ulem}
\usepackage{cancel}

\begin{document}

\title[Spin thermopower in the overscreened Kondo model]{Spin thermopower
in the overscreened Kondo model}

\author{R \v{Z}itko $^{1,2}$, J. Mravlje $^{1,3}$, A. Ram\v{s}ak $^{1,2}$, T. Rejec $^{1,2}$}
\address{$^1$ J. Stefan Institute, Jamova 39, SI-1000 Ljubljana, Slovenia}
\address{$^2$ Faculty for Mathematics and Physics, University of
Ljubljana, Jadranska 19, SI-1000 Ljubljana, Slovenia}
\address{$^3$ Coll\`ege de France, 11 place Marcelin Berthelot, 75005
Paris, France}
\ead{rok.zitko@ijs.si}

\date{\today}

\begin{abstract}
We study the spin thermopower in the two-channel spin-$1/2$ Kondo
model which exhibits the phenomenon of impurity spin overscreening and 
non-Fermi-liquid properties. While magnetic field lower than the Kondo
temperature does not strongly polarize the impurity spin, we show that it
nevertheless strongly affects the low-energy part of the spectral
function. In turn, this leads to characteristic saturation of the spin
Seebeck coefficient at the value of $0.388 k_B/|e|$ at $T \sim T^*$,
where $T^* \propto B^2/T_K$ is the scale of the crossover between the
intermediate-temperature non-Fermi-liquid regime and the
low-temperature Fermi-liquid regime. We show that measuring the
spin thermopower at low magnetic fields would provide a sensitive test for
distinguishing regular Fermi liquid, singular Fermi liguid, and
non-Fermi liquid behaviour in nanodevices.
\end{abstract}

\pacs{72.10.Fk, 72.15.Qm}

\maketitle

\newcommand{\vc}[1]{{\mathbf{#1}}}
\newcommand{\ket}[1]{|#1\rangle}
\newcommand{\bra}[1]{\langle #1|}
\newcommand{\braket}[1]{\langle #1 \rangle}
\renewcommand{\Im}{\mathrm{Im}}
\renewcommand{\Re}{\mathrm{Re}}
\newcommand{\dr}{\mathrm{d}}
\newcommand{\correl}[1]{\langle\langle #1 \rangle\rangle_\omega}
\newcommand{\TKO}{T_K^{(0)}}
\newcommand{\TKt}{T_K^{(2)}}

\newcommand{\figw}{7cm}
\newcommand{\dIdV}{\mathrm{d}I/\mathrm{d}V}

\bibliographystyle{unsrt}

\section{Introduction}

The Kondo effect is one of the most intensely studied and
well-understood many-particle effects which occur due to strong
interactions \cite{hewson}. It may be adequately described by the
Kondo model, a simple s-d exchange Hamiltonian for a point-like
spin-$1/2$ impurity coupled to a continuum of conduction-band
electrons through an antiferromagnetic exchange interaction. Due to
strong confinement of electrons in nanostructures (such as
semiconductor quantum dots \cite{goldhabergordon1998b}, atoms and
molecules \cite{roch2009}, fullerenes \cite{yu2004a}, segments of
carbon nanotubes \cite{nygard2000}, etc.), the Kondo effect is fairly
ubiqitous in various nanostructures exhibiting the Coulomb blockade
phenomenon \cite{kouwenhoven2001}. It manifests as enhanced
conductivity in valleys with odd electron occupancy when the
temperature is reduced below the characteristic Kondo temperature,
$T_K$. Many experiments have confirmed the universal properties
predicted for the Kondo effect and demonstrated that simple impurity
models well describe most experimental features.
At low temperatures, most such models behave as regular Fermi liquids
(RFL) \cite{nozieres1974}: the low-energy excitations are conventional
fermionic quasiparticles \cite{hewson2004,hewson2005} which have
analytical scattering properties in the vicinity of the Fermi energy
\cite{mehta2005}.

In recent years, experimental realizations of more exotic types of the
Kondo effect have also been reported. The underscreened Kondo effect
occurs when a spin-$1$ impurity is effectively coupled to a single
conduction band \cite{roch2009,parks2010,logan2009,vzporedneqmc}. The impurity spin is only
partially screened and there is residual impurity magnetization and
residual impurity entropy
\cite{cragg1979b,sacramento1989,coleman2003,koller2005}.  The
scattering amplitude of electrons near the Fermi energy is not
analytical and the conventional Fermi-liquid picture breaks down
\cite{mehta2005,koller2005}. Such systems are classified as singular
Fermi liquids (SFL) \cite{mehta2005}. Experimentally, this type of
behaviour is found in molecules with a spin center, which are
stretched between two electrodes in mechanically controllable break
junctions \cite{roch2009,parks2010}. In these systems, the spin would
in principle be fully screened, but the Kondo temperature of the
second screening stage is much below the experimental temperature. 

Another unusual type of the Kondo effect, the overscreened Kondo
effect, occurs when a spin-$1/2$ impurity is effectively coupled to
two independent conduction bands. The impurity is overscreened and an
unusual non-Fermi-liquid (NFL) ground state emerges
\cite{nozieres1980,affleck1991over,affleck2005,toth2008prb}. The
low-energy excitations are not Dirac-electron-like quasiparticles, but
rather Majorana fermions \cite{maldacena1997}. Experimentally, such
behaviour was observed in an experiment where a small quantum dot (the
impurity) was coupled to conduction leads (first conduction band) and
a big quantum dot (second conduction band) \cite{potok2007}. By
careful tuning of system parameters, it has been shown that the
transport properties scale with temperature and gate voltage in
accordance with the two-channel Kondo (2CK) model.

The characterization of nanodevices through transport properties
provides information about the spectral function of the system. The
low-temperature differential conductance is a measure of the spectral
density at the Fermi level. As it is relatively easy to measure, it
remains the most commonly used experimental probe. Additional
information can be gathered by measuring other transport properties,
such as the thermopower (Seebeck coefficient), which is a sensitive
probe of the particle-hole asymmetry of the spectral function
\cite{costi2010,cornaglia2012}. In this work, we study a related
property, the spin thermopower (spin Seebeck coefficient)
\cite{rejec2012,cornaglia2012}. It is defined as the induced spin
voltage across the device at zero spin current in the presence of a
temperature gradient. The spin Seebeck coefficient $S_s$ is a
sensitive probe of the spectral peak splitting in the presence of
external magnetic field. Since the different kinds of the Kondo effect
have very different response to an applied magnetic field, they can be
clearly distinguished by different behaviour of $S_s$ as a function of
the magnetic field $B$ and the temperature $T$, as will become evident
in the following.

The spin thermopower in the conventional Kondo model (single-channel
$S=1/2$ model, which is a RFL) has been studied in
Ref.~\cite{rejec2012}. For $T \lesssim T_K$, the spin thermopower as a
function of $B$ peaks at $B \sim T_K$ ($B$ is measured in the units of
the Zeeman energy $g\mu_B B$, while the temperature is measured in the
units of energy $k_B T$). For $B \lesssim T_K$, the spin thermopower
as a function of $T$ peaks at $T \sim 0.3 T_K$. Finally, for $B, T
\gtrsim T_K$, the spin thermopower peaks at $B \sim 3 T$. These
features can be well understood within a simple model for the Kondo
peak splitting in applied magnetic field, taking into account thermal
broadening effects. 

The spin thermopower in the underscreened Kondo model (single-channel
$S=1$ model, which is a SFL \cite{coleman2003,koller2005}) has been
studied in Ref.~\cite{cornaglia2012}. Due to the decoupled residual
impurity moment, even a tiny magnetic field strongly polarizes the
impurity and splits the Kondo peak. The main difference between
the $S=1/2$ and $S=1$ models is thus in the $B \lesssim T_K$ regime, where the
maximum in the spin thermopower as a function of $T$ is now found for
$T \sim 0.2 B$, rather than at constant $T \sim 0.3 T_K$; the
amplitude of the maximum is comparable in both cases, being somewhat
smaller in the underscreened case in a wide parameter range.

The spin thermopower in the overscreened Kondo model (two-channel
$S=1/2$ Kondo model, which is a true NFL system) is the subject of
this work. The magnetic field is a relevant perturbation which drives
the system away from the NFL fixed point, which governs the behaviour
of the system in the temperature range $T^* \ll T \ll T_K$, to the
stable low-temperature RFL fixed point. We will show that the 2CK
model exhibits very peculiar low-field behaviour of the spin
thermopower, which has a maximum at the crossover scale $T^* \propto
B^2/T_K$ with a height which {\it saturates} at small $B$ (unlike in
the fully screened and underscreened cases, where the height goes to
zero). The saturated value at the maximum is $0.388 k_B/|e| \approx
\unit[34]{\mu V/K}$. We will compare the very interesting contrasting
behaviour of the underscreened and overscreened models: while in the
underscreened model, a small magnetic field that strongly polarizes
the impurity spin affects mainly the high-energy part of the spectral
function, in the overscreened model the polarization effect is small
but it strongly affects especially the low-energy part of the spectral
function, thus leading to the saturation in the spin Seebeck
coefficient.

\section{Models, spin thermopower, method}
\label{sec2}

\begin{figure}
\centering
\includegraphics[width=13cm]{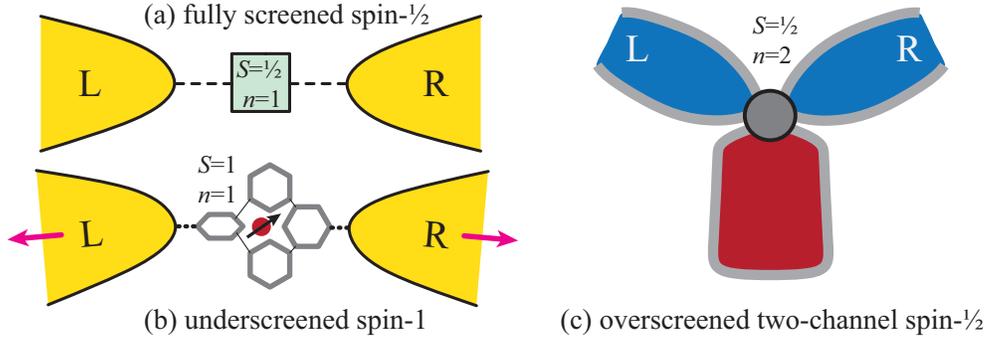}
\caption{Schematic representation of the three nanodevices which
can be described using the spin-$S$ $n$-channel Kondo model. (a)
Semiconductor quantum dot. (b) Spin-1 molecule \cite{roch2009,parks2010}.
(c) Quantum dot coupled to a large quantum dot \cite{potok2007}.}
\label{skica}
\end{figure}

The physical systems discussed in this work are schematically
represented in Fig.~\ref{skica}. The nanodevice is coupled to two
contacts kept at different temperatures. The electrons in these two
contacts form one screening channel. In addition, in the 2CK case, the
role of the second screening channel is played by the electrons in the
additional large quantum dot. The electrons in this dot are
electrostatically confined by the large Coulomb repulsion in order to
eliminate processes in which an electron would tunnel from the large
quantum dot to the contacts; such processes would destroy the
two-channel physics (i.e., they correspond to an additional relevant
operator which drives the system away from the NFL fixed point).

At low enough temperatures, such nanosystems can all be modeled using
some variant of the Kondo impurity model, where the impurity degrees
of freedom are described using a quantum mechanical spin operator
$\vc{S}$ which is locally coupled with the spin density of the
conduction band electrons. The corresponding Hamiltonian is thus
\begin{equation}
H=\sum_{\vc{k},\sigma,i} \epsilon_k c^\dag_{\vc{k},\sigma,i}
c_{\vc{k},\sigma,i}
+ \sum_\alpha J \vc{s}_i \cdot \vc{S} + \vc{B} \cdot \vc{S},
\end{equation}
where $c^\dag_{k,\sigma,i}$ are creation operators for the itinerant
electrons in channel $i$ with momentum $\vc{k}$, spin $\sigma$
and energy $\epsilon_k$. The spin density of the band $i$ can be
expressed as
\begin{equation}
\vc{s}_i = \frac{1}{N} \sum_{\vc{k},\vc{k}'} \sum_{\alpha\beta}
c^\dag_{\vc{k},\alpha,i} \left( \frac{1}{2}
\boldsymbol{\sigma}_{\alpha\beta} \right)
c_{\vc{k}',\beta,i} 
= \sum_{\alpha\beta} f^\dag_{0,\alpha,i}  \left( \frac{1}{2}
\boldsymbol{\sigma}_{\alpha\beta} \right)
f_{0,\beta,i}.
\end{equation}
Here $f^\dag_{0,\sigma,i} = (1/\sqrt{N}) \sum_{\vc{k}}
c^\dag_{\vc{k},\sigma,i}$ is the combination of states which couples
to the impurity, while $\boldsymbol{\sigma}=\{
\sigma^x, \sigma^y, \sigma^z
\}$ is a vector of Pauli matrices. The index $i$ ranges over the
number of channels $n$ (one or two). Depending on the value of the
impurity spin $S$ and the number of channels $n$, the impurity spin
will be either fully screened ($n=2S$), underscreened ($n<2S$) or
overscreened ($n>2S$)
\cite{nozieres1980,furuya1982,andrei1984,sacramento1991td,zarand2004,borda2007inelastic}.
We fix the Kondo exchange coupling $J$ to a constant value $J/D=0.2$.
Here $D$ is the half-width of the conduction band which is assumed to
have a flat density of states $\rho=1/2D$. The half-bandwidth is also
used as the energy unit, $D=1$. 
The Kondo temperature is the same (up
to non-exponential prefactors) in all three models, $T_K \propto
\exp(-1/\rho J)$. Numerically, we find
\begin{equation}
T_{K,W}=1.1\times 10^{-5},
\end{equation}
where $T_{K,W}$ is the Kondo temperature as defined by Wilson through
the impurity magnetic susceptibility,
$T_{K,W}\chi_\mathrm{imp}(T_{K,W})=0.07$
\cite{wilson1975,krishna1980a}. The magnetic field is measured in
units of the Zeeman energy (i.e., we absorb the factor $g\mu_B$ into
$B$).

In this work we focus on the spin Seebeck coefficient $S_s$, defined
as \cite{rejec2012}
\begin{equation*}
S_s = -\frac{eV_s}{\Delta T}\Bigr|_{I_s=0}.
\end{equation*}
The spin thermopower is the ability of the device to convert
the temperature difference $\Delta T$ to spin voltage $V_s$, assuming no
spin current $I_s$ flows through the impurity. In units of $k_B/|e|$,
to be used in what follows, the spin Seebeck
coefficient
\begin{equation}
S_s = \frac{2}{T}\frac{\mathcal{I}_1}{\mathcal{I}_0}
\end{equation}
is determined by computing the transport integrals
\begin{equation}
\label{integral}
\mathcal{I}_{n\sigma} = \int \mathrm{d}\omega\, \omega^n
[-f'(\omega)] \mathcal{T}_\sigma(\omega),
\end{equation}
where $f(\omega)$ is the Fermi-Dirac function at temperature $T$. At
the particle-hole symmetric point (the case considered in this work),
one has $\mathcal{I}_{0\uparrow}=\mathcal{I}_{0\downarrow} \equiv
\mathcal{I}_0$ and $\mathcal{I}_{1\uparrow}=-\mathcal{I}_{1\downarrow}
\equiv \mathcal{I}_1$. From the definition it is manifest that the
spin thermopower effect probes the difference in the particle-hole
asymmetry of the two spin species.

The quantity $\mathcal{T}_\sigma$ in Eq.~(\ref{integral}) is the
transmission function of electrons with spin $\sigma$ in the first
channel (i.e., the channel which corresponds to the source and drain
leads). It is proportional to the imaginary part of the $T$ matrix for
the Kondo model, defined as $G=G_0+G_0TG_0$, where $G$ is the
interacting Green's function for electrons, while $G_0$ is the
non-interacting one. For a given model, it may be computed using the
equation of motion approach. One finds $T_\sigma = \langle\langle
O_\sigma; O_\sigma^\dag \rangle\rangle$, where $O_\sigma$ is defined
as the commutator between the annihilation operator on the first site
of the Wilson chain (which is the same as the combination of
conduction-band states which couples to the impurity,
$f_{0,\sigma,1}$) and the Kondo coupling term,
$O_\sigma=[H_{K,1},f_{0,\sigma,1}]$, where $H_{K,1} = J \vc{s}_1 \cdot
\vc{S}$. The result is a composite fermion operator
\cite{costi2000,rosch2003}
\begin{equation}
O_\alpha = \left( \frac{1}{2}
\sum_\beta \boldsymbol{\sigma}_{\alpha\beta}
f_{0,\beta,1} \right) \cdot \vc{S}.
\end{equation}

The transport integrals have been computed using the numerical
renormalization group (NRG) method
\cite{wilson1975,krishna1980a,bulla2008}, using the approach described
in Ref.~\cite{rejec2012}: the Lehmann representation of the spectral
function is integrated over and the transport integrals
$\mathcal{I}_{n\sigma}$ computed for all temperatures in a single NRG
sweep. It has been shown that for the RFL case this
leads to only a very small deviation (few percent at most) from the
results of the more accurate full-density-matrix (FDM NRG) approach
\cite{weichselbaum2007}, and we have now verified that this is the
case also for the SFL in magnetic field despite the
strong magnetic polarizability of this model.

We conclude this section by a remark about the role of the second
quantum dot as regards the particle and heat flow. We have assumed
that the voltage and the temperature in the leads are symmetrically
shifted with respect to the equilibrium value. For this reason,
neither charge nor spin current flow to the large quantum dot. The
large dot thus only plays the role of an additional screening channel.

\section{Spin Seebeck coefficient of the overscreened Kondo effect}
\label{sec3}

\begin{figure}
\centering
\includegraphics[width=13cm]{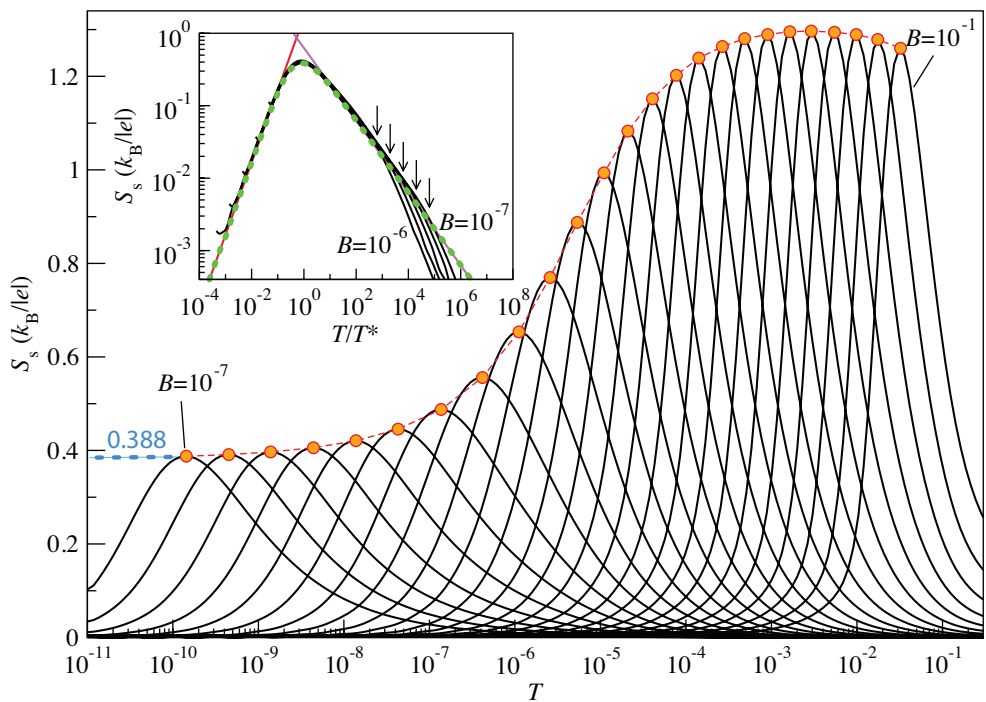}
\caption{Spin Seebeck coefficient for the two-channel Kondo model. The
magnetic field ranges from $B=10^{-7}$ to $B=10^{-1}$ in steps of
factor $10^{1/4}$.  The colored dots indicate the positions of the
peaks in the spin Seebeck coefficient for different values of $B$; the
positions, $T_\mathrm{max}$, and the corresponding values of the
coefficient, $S_s(T_\mathrm{max})$, are plotted in Fig.~\ref{figcomp}
as a function of $B$.
The inset shows how, for temperatures below the Kondo
temperature (indicated by arrows), the curves for different fields
from $B=10^{-7}$ to $B=10^{-6}$ collapse into a universal curve
given by Eq.~(\ref{eq:ss}) and shown with green dotted line. The
low temperature, $T\ll T^{\ast}$, asymptote of Eq.~(\ref{eq:sslow})
is shown as red line. The intermediate temperature, $T^{\ast}\ll T\ll
T_{K}$, asymptote of Eq.~(\ref{eq:sshigh}) is shown as violet line.}
\label{fig0}
\end{figure}

The main result of this work is shown in Fig.~\ref{fig0}. We plot the
spin Seebeck coefficient of the 2CK model for a range of applied
external magnetic fields $B$ as a function of $T$.

For any $B$, the spin thermopower curve has a single peak. For $B \ll
T_K$, the peak occurs at the NFL-RFL crossover scale $T^* \propto
B^2/T_K$. The $B^2$ scaling is a characteristic feature of the model
and is not expected to be observed in single-channel models.  In this
regime, $S_s(T)$ at the point of its maximum saturates at a value that
is determined using analytical arguments in the following section.  (A
somewhat related saturation of the charge Seebeck coefficient has been
noted in a different proposed physical realization of the 2CK model
\cite{nguyen2010}. In that case, the symmetry breaking operator was
the channel asymmetry, $J_1 \neq J_2$.) The behaviour is universal and
we find that the curves, when plotted as as a function of $T/T^*$,
fully overlap at low temperatures, see the inset in Fig.~\ref{fig0}.

For $B \gg T_K$, the spin thermopower peaks on the temprature scale of
$B$. Such asymptotic behaviour is expected for any impurity model
where the local degrees of freedom are described by a spin operator.
The reason is that for large $B$, there will be a threshold for
inelastic spin excitations given by the Zeeman energy. The spectral
lineshapes, however, are expected to be model dependent.
(The small drop in the maximum of $S_s(T)$ at the largest values
of $B$ in Fig.~\ref{fig0} is due to finite-bandwidth effects.)

\section{Analysis of the asymptotic behaviour and the NFL-RFL crossover}
\label{sec4}

The NFL state of the 2CK model is destabilized by a perturbation due
to an external magnetic field acting on the impurity, forcing the
system towards a RFL ground state. In Ref.~\cite{Mitchell12} the
authors exploited a connection to an exactly solvable classical
boundary Ising model to obtain the exact finite-temperature
transmission function,
\begin{eqnarray*}
{\cal T}_{\sigma}\left(\omega,T\right) & = & \frac{1}{2}+\sigma\frac{1}{\sqrt{8\pi^{3}}}\int_{-\infty}^{\infty}\mathrm{d}x\frac{\cos\frac{x\omega}{\pi T}}{\tanh\frac{\omega}{2T}\sinh x}\times\\
 &  & \times\mathrm{Re}\left\{ \sqrt{\frac{T^{\ast}}{T}}\frac{\Gamma\left(\frac{1}{2}+\frac{1}{2\pi}\frac{T^{\ast}}{T}\right)}{\Gamma\left(1+\frac{1}{2\pi}\frac{T^{\ast}}{T}\right)}\,_{2}F_{1}\left(\frac{1}{2},\frac{1}{2},1+\frac{1}{2\pi}\frac{T^{\ast}}{T},\frac{1-\coth x}{2}\right)\right\} ,
\end{eqnarray*}
describing the crossover from low temperature, $T\ll T^{\ast}$,
FL to intermediate temperature, $T^{\ast}\ll T\ll T_{K}$, NFL
regime. Here $\Gamma$ is the Gamma function and $_{2}F_{1}$
is the hypergeometric function. The crossover temperature scale,
\[
T^{\ast}=c_{B}^{2}\frac{B^{2}}{T_{K}},
\]
is proportional to the square of the magnetic field, which should be
small enough, $B\ll T_{K}$, to ensure that the crossover temperature
is well separated from the Kondo temperature. $c_{B}={\cal
O}\left(1\right)$ is a fitting parameter, in our case $c_{B}=0.436$.
This theory applies exactly in the $B \ll T_K$ limit. We find that for
the range of $B$ considered in this work, the transmission function
has a multiplicative field-dependent prefactor, which however largely
cancels out when $S_s(T)$ is computed as a ratio of two transport
integrals, thus the universal scaling for $S_s(T)$ applies well for
$B/T_K \lesssim 10^{-1}$.

To evaluate the spin Seebeck coefficient, the integrals over $\omega$
in Eqs.~(3) and (4) can be calculated exactly. The spin Seebeck coefficient,
\begin{eqnarray}
\label{num1}
S_{s} & = & \frac{\sqrt{\pi}}{2\sqrt{2}}\int_{-\infty}^{\infty}\mathrm{d}x\frac{1}{\cosh^{2}\frac{x}{2}\sinh x}\times\nonumber \\
 &  & \times\mathrm{Re}\left\{ \sqrt{\frac{T^{\ast}}{T}}\frac{\Gamma\left(\frac{1}{2}+\frac{1}{2\pi}\frac{T^{\ast}}{T}\right)}{\Gamma\left(1+\frac{1}{2\pi}\frac{T^{\ast}}{T}\right)}\,_{2}F_{1}\left(\frac{1}{2},\frac{1}{2},1+\frac{1}{2\pi}\frac{T^{\ast}}{T},\frac{1-\coth x}{2}\right)\right\} ,\label{eq:ss}
\end{eqnarray}
is a universal function of rescaled temperature $T/T^{\ast}$. The
inset to Fig.~1 shows how the NRG curves corresponding to different
magnetic fields $B\ll T_{K}$ collapse into a universal curve at
temperatures well below $T_{K}$. In this regime, the spin Seebeck
coefficient reaches its maximum value of
$S_{s}(T_\mathrm{max})=0.388$, independently of the field, at
$T_\mathrm{max}=0.829T^{\ast}$. We were unable to find a closed-form expression
for $S_{s}(T_\mathrm{max})$, but it can be computed numerically as the
maximum value of the expression in Eq.~(\ref{num1}).

At low temperatures, $T\ll T^{\ast}$, the transmission function is well approximated
by its zero temperature form. For $\omega\ll T^{\ast}$ the
transmission function
deviates form its $\omega=0$ value, ${\cal T}_{\sigma}\left(0,0\right)=\frac{1}{2}$,
in a linear fashion,
\begin{equation}
{\cal T}_{\sigma}\left(\omega,0\right)\overset{\,\,\,\omega\ll T^{\ast}}{\sim}\frac{1}{2}+\frac{\sigma}{8}\frac{\omega}{T^{\ast}}.\label{eq:tmatlow}
\end{equation}
The deviation $\left|{\cal T}_{\sigma}\left(\omega,0\right)-\frac{1}{2}\right|$
reaches its maximal value of 0.13 at $\omega\approx4T^{\ast}$ and
drops to zero as 
\[
{\cal T}_{\sigma}\left(\omega,0\right)\overset{\,\,\,\omega\gg T^{\ast}}{\sim}\frac{1}{2}+\frac{\sigma}{2\sqrt{2}\pi}\frac{\ln\omega}{\sqrt{\omega}}
\]
in the high frequency limit. The low temperature spin Seebeck coefficient,
probing the low frequency part of the transmission function, Eq.~(\ref{eq:tmatlow}),
\begin{equation}
S_{s}\overset{\,\,\, T\ll
T^{\ast}}{\sim}\frac{2\pi^{2}}{3}T\left.\frac{\partial\ln {\cal T}_{\uparrow}\left(\omega,0\right)}{\partial\omega}\right|_{\omega=0}=\frac{\pi^{2}}{6}\frac{T}{T^{\ast}},\label{eq:sslow}
\end{equation}
exhibits a typical Fermi liquid linear dependence on temperature.

At higher temperatures, $T^{\ast}\ll T\ll T_{K}$, the transmission
function takes a simpler form, 
\begin{eqnarray*}
{\cal T}_{\sigma}\left(\omega,T\right) & = &
\frac{1}{2}+\sigma\sqrt{\frac{T^{\ast}}{T}}\times \\
 & & \times\frac{1}{\sqrt{2\pi^{4}}}\int_{-\infty}^{\infty}\mathrm{d}x\frac{\cos\frac{x\omega}{\pi T}}{\tanh\frac{\omega}{2T}\sinh x}\mathrm{Re}\left\{K\left(\frac{1-\coth x}{2}\right)\right\} ,
\end{eqnarray*}
where $K$ is the complete elliptic integral of the first kind. It
can be cast in a scaling form, 
\[
\frac{{\cal T}_{\sigma}\left(\omega,T\right)-\frac{1}{2}}{\sqrt{\frac{T^{\ast}}{T}}}=\sigma{\cal F}\left(\frac{\omega}{T}\right),
\]
where the scaling function ${\cal F}$ is linear in frequency for
$\omega\ll T$, 
\[
{\cal F}\left(u\right)\overset{u\ll1}{\sim}0.082u,
\]
reaches a maximum value of 0.15 at $\omega\approx5T$ and drops to zero as
\[
{\cal F}\left(u\right)\overset{u\gg1}{\sim}\frac{1}{2\sqrt{2}\pi}\frac{\ln u}{\sqrt{u}}
\]
in the high frequency limit. Such a scaling causes the spin Seebeck
coefficient to exhibit an anomalous NFL dependence on temperature
\begin{equation}
S_{s}\overset{\,\,\, T\gg T^{\ast}}{\sim}\frac{4\sqrt{2}}{9}\sqrt{\frac{T^{\ast}}{T}}.
\label{eq:sshigh}
\end{equation}

To sum it up, the spin Seebeck coefficient of the 2CK model at low $B$
is characterized by an asymmetric peak of a universal form with a
maximum value of about 0.4 at the crossover temperature $T^{\ast}$
between RFL and NFL regimes. The low temperature slope is proportional
to $T/T^{\ast}$, while the high temperature slope is proportional to
$\sqrt{T^{\ast}/T}$.

\section{Contrasting behaviour of spin thermopower in regular, singular
and non-Fermi liquids}
\label{sec5}

\begin{figure}
\centering
\includegraphics[width=17cm]{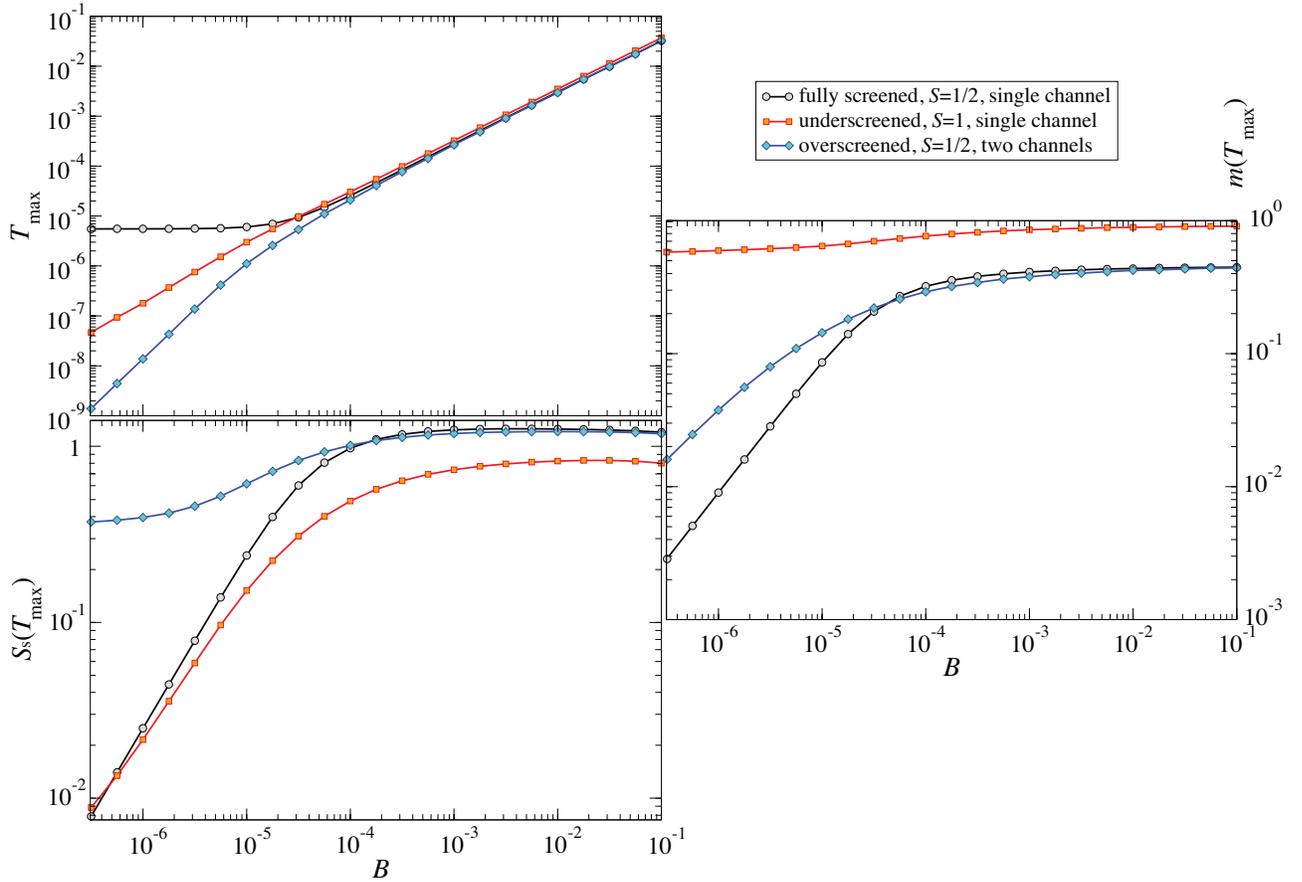}
\caption{Top left: position of the peak of the spin Seebeck coefficient as
a function of $T$ for various values of the magnetic field $B$.
Bottom left: value of the spin Seebeck coefficients at its peak.  Right:
the impurity spin polarization as a function of the magnetic field $B$
at the temperature of the peak in the spin Seebeck coefficient.}
\label{figcomp}
\end{figure}

In Fig.~\ref{figcomp} we compare the main characteristics of the spin
thermopower in the three models considered. As a function of the
applied magnetic field we plot the following quantities: the temperature
$T_\mathrm{max}$ of the maximum of the spin Seebeck coefficient $S_s(T)$,
the value $S_s(T_\mathrm{max})$ of the spin Seebeck coefficient at this maximum,
and the impurity magnetic moment $m(T_\mathrm{max})=\langle S_z \rangle$ at the maximum of $S_s$.
The first two quantities have been
extracted from the results shown in Figs.~\ref{fig0} and \ref{fig0b}.

We first note that the high-field behaviour is rather similar in all
three models, as has been already suggested in Sec.~\ref{sec3}.  We
find that the temperature of the maximum is almost the same in all
three cases and given approximately by $T_\mathrm{max} \approx 0.3 B$.
The value of spin Seebeck coefficient at this maximum depends only on
the impurity spin, not on the number of channels, and it saturates for
high magnetic fields: for $S=1/2$ we find
$S_s(T_\mathrm{max}) \approx 1.32$,
while for $S=1$ we obtain a {\it smaller} value
$S_s(T_\mathrm{max}) \approx 0.9$.
The magnetizations at the maximum are also spin dependent:
for $S=1/2$ we find $m(T_\mathrm{max}) \approx 0.45$,
while for $S=1$ it is twice as large,
$m(T_\mathrm{max}) \approx 0.9$.

A crude way to understand the high-field asymptotic behaviour is to
postulate the following rough approximation for the transmission
function in the large-$B$ limit:
\begin{equation}
\label{crude}
\mathcal{T}_\sigma(\omega) \propto 1 + c m \theta(\sigma \omega-B),
\end{equation}
i.e., a piecewise constant function with a jump at $\omega=\sigma B$
by a value proportional to the magnetization with $c$ a constant of
order 1. The increase in scattering is due to the opening of the
inelastic scattering channel due to spin-flips, which is possible for
energies above the Zeeman energy. This form is suggested by the more
accurate renormalization group theory \cite{rosch2003}. We find that
the spin Seebeck coefficient obtained using Eq.~(\ref{crude}) has a
peak at $T=2.3 B$ with a height of $0.89$ (for $c=1$ and taking the
value of $m$ from NRG). Considering the extreme crudeness of the
model, the result is a fair approximation for the true values.

For all three models, the crossover to the low-field behaviour occurs
on the same scale of the Kondo temperature. The low-field asymptotic
behaviour, however, is markedly different in the three models. In the
RFL case, the maximum occurs at the constant temperature of $T \sim
0.3 T_K$ \cite{rejec2012}.  This is because the impurity spectral
function is not significantly affected by magnetic fields lower than
$T_K$, there is only a rigid shift of the two spin-projected
components of the spectral function. The linear reduction of
$S_s(T_\mathrm{max})$ is also well accounted for by simple Fermi
liquid arguments \cite{rejec2012}.
RFLs exhibit Pauli paramagnetism with $T_K$ playing the role of the Fermi
temperature, thus the magnetization of this model at low temperatures
is given by $B/T_K$.

In the SFL, the spin Seebeck coefficient peaks at a temperature on the
scale where the system crosses over from the SFL regime at
intermediate temperatures to the asymptotic RFL behaviour at low
temperatures \cite{mehta2005}. The crossover temperature, defined from
the thermodynamic properties as the temperature where the impurity
entropy decreases to one half of its value at the SFL fixed point, is
found to be $T_\mathrm{td} = 0.32 B$, with no logarithmic correction.
The position of the peak in $S_s$, $T_\mathrm{max}$, appears to have
logarithmic corrections compared to $T_\mathrm{td}$. The maximum spin
Seebeck coefficient $S_s(T_\mathrm{max})$ decreases to zero as a
linear function, but again with logarithmic corrections.

For completeness we also plot the impurity spin magnetization
for the three models in the right panel of Fig.~\ref{figcomp}.
We emphasize that strong spin polarization does not
necessarily imply a large spin Seebeck coefficient.
In fact, the underscreened model with by far the largest polarization
has the smallest spin Seebeck coefficient in a wide interval
of magnetic fields (in fact, only for $B < 10^{-2}T_K$ is
its spin Seebeck coefficient somewhat larger than that of the fully
screened model).

\begin{figure}
\centering
\includegraphics[width=10cm]{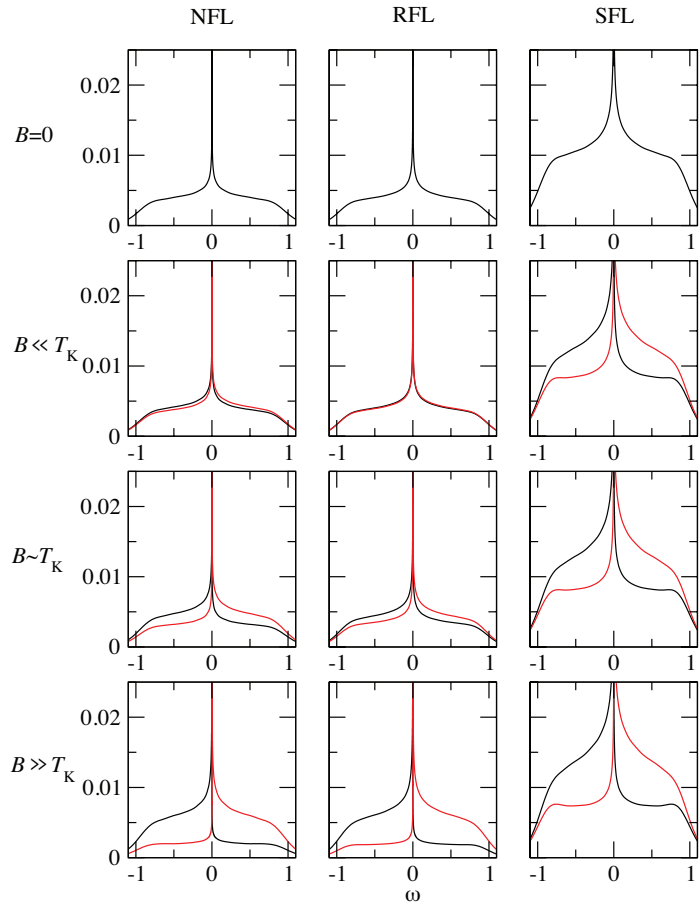}
\caption{$T=0$ spectra of the three models for a range of
magnetic fields. The magnetic fields are $B =10^{-6} \ll T_K$,
$B=10^{-5} \sim T_K$ and $B=10^{-4} \gg T_K$.
Red curves: spin up, $\sigma=\uparrow$, black curves: spin down,
$\sigma=\downarrow$.
} \label{figall}
\end{figure}

\begin{figure}
\centering
\includegraphics[width=10cm]{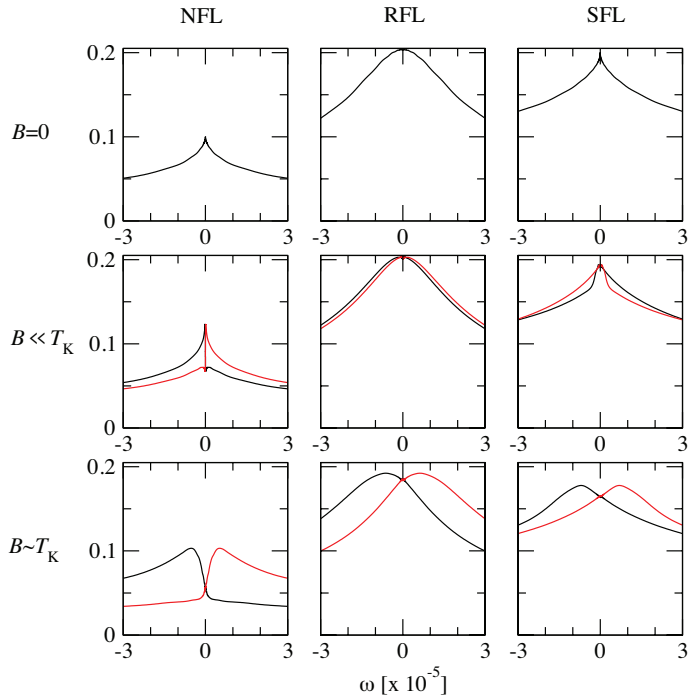}
\caption{$T=0$ spectra of the three models for a range of
magnetic fields: close-up on the low-frequency region.}
\label{figzoom}
\end{figure}

To better understand the value of the spin Seebeck coefficient we
study the impurity spectral functions, Fig.~\ref{figall}. At $B=0$,
the overall aspect of the spectral functions is the same: they all
have a Kondo resonance with a width proportional to $T_K$. The main
differences are better visible in the close-up to the peak region,
Fig.~\ref{figzoom}. The resonance shape is analytical only in the RFL
case. The approach to the asymptotic value in the SFL is logarithmic
\cite{koller2005}, while in the 2CK effect it has a square root
non-analyticity \cite{affleck1993,toth2007}. Note also that the
$\omega=0$ value of the spectral function is twice as big in RFL and
SFL as compared to the NFL case \cite{toth2007}. For small fields, $B
\ll T_K$, the effect on the overall spectral function is the most
significant in the SFL case where a strong spin polarization clearly
manifests on all frequency scales, see Fig.~\ref{figall}. The close-up
in the low-frequency region, Fig.~\ref{figzoom}, however reveals that
in the NFL case the relative change of the spectral function is very
significant. The spin-up spectral function is much higher for
$\omega>0$ compared to $\omega<0$ (by almost a factor of 2), and it
peaks on the frequency scale of $T^* \propto B^2/T_K$. The peak height
even exceeds the height of the zero-field spectral function at
$\omega=0$. This low-frequency polarization effect is much stronger
compared to that for (singular and regular) Fermi liquid systems. This
explains the large spin Seebeck coefficient in the NFL case.

\section{Conclusion}

We have analyzed the spin thermopower of the two-channel Kondo model,
which is a paradigmatic case of an impurity model with
non-Fermi-liquid properties at low energy scales. We have shown that
the crossover to the Fermi-liquid ground state generated by applied
magnetic field $B < T_K$ leads to a pronounced peak in the spin
Seebeck coefficient as a function of the temperature. The peak
position correspond to the crossover scale $T^* \propto B^2/T_K$,
while its height saturates at a sizeable value of $0.388 k_B/|e|$. This
is to be opposed with the single-channel Kondo models with $S=1/2$
(RFL) and $S=1$ (SFL), where for $B < T_K$ the peak height decreases
at low $B$. We have shown that this behaviour is due to a
characteristic effect of the magnetic field on the spectral function,
which in the 2CK model becomes strongly spin-dependent
on the lowest energy scales. This is in stark contrast with the
strongly spin polarizable underscreened $S=1$ model, where the
spin-dependence is large on high energy scales, but is less pronounced
in the region $-5k_B T<\omega<5k_BT$ which is relevant for transport.
An analytical approach can be applied to the universal low-temperature
regime of the two-channel Kondo model and we have shown that the
saturation value of $\sim 0.388 k_B/|e|$ can be computed numerically to
high precision.

The most important outcome of this work is, perhaps, the observation
of very different behaviour of the spin thermopower in the three
classes of quantum impurity models. We speculate that all models with
$S=n/2$ (full screening) are characterized by a spin Seebeck
coefficient peak at $\sim T_K$ with decreasing height, all models with
$S>n/2$ (underscreening) are characterized by a peak at $\sim B$
(perhaps with some logarithmic corrections) with decreasing height,
while in models with $S<n/2$ (overscreening) the spin Seebeck
coefficient peaks at the crossover scale and its height saturates.
Here $S$ is the impurity spin, while $n$ is the number of screening
channels.

On the applied side, the spin-thermoelectric effect might be of
interest for spintronics as it enables generation of the spin
current without any accompanying charge current. In this respect we
note that the the overscreened Kondo model outperforms the exactly
screened and the underscreened ones, especially at magnetic fields
below and around the Kondo temperature. The universal dependence of spin
Seebeck coefficient on temperature with the peak position specified by
magnetic field could be considered also for sensitive measurements of
the magnetic field.

\appendix

\section{Spin Seebeck coefficient in single-channel problems}

For reference and easier comparisons, in this appendix we provide the
results for the spin Seebeck coefficient for the single-channel Kondo
models.  They are computed in exactly the same way and with exactly
the same parameters as in Fig.~\ref{fig0}. The positions
of the maxima in the spin Seebeck coefficient are plotted in
Fig.~\ref{figcomp}.

\begin{figure}
\centering
\includegraphics[width=13cm]{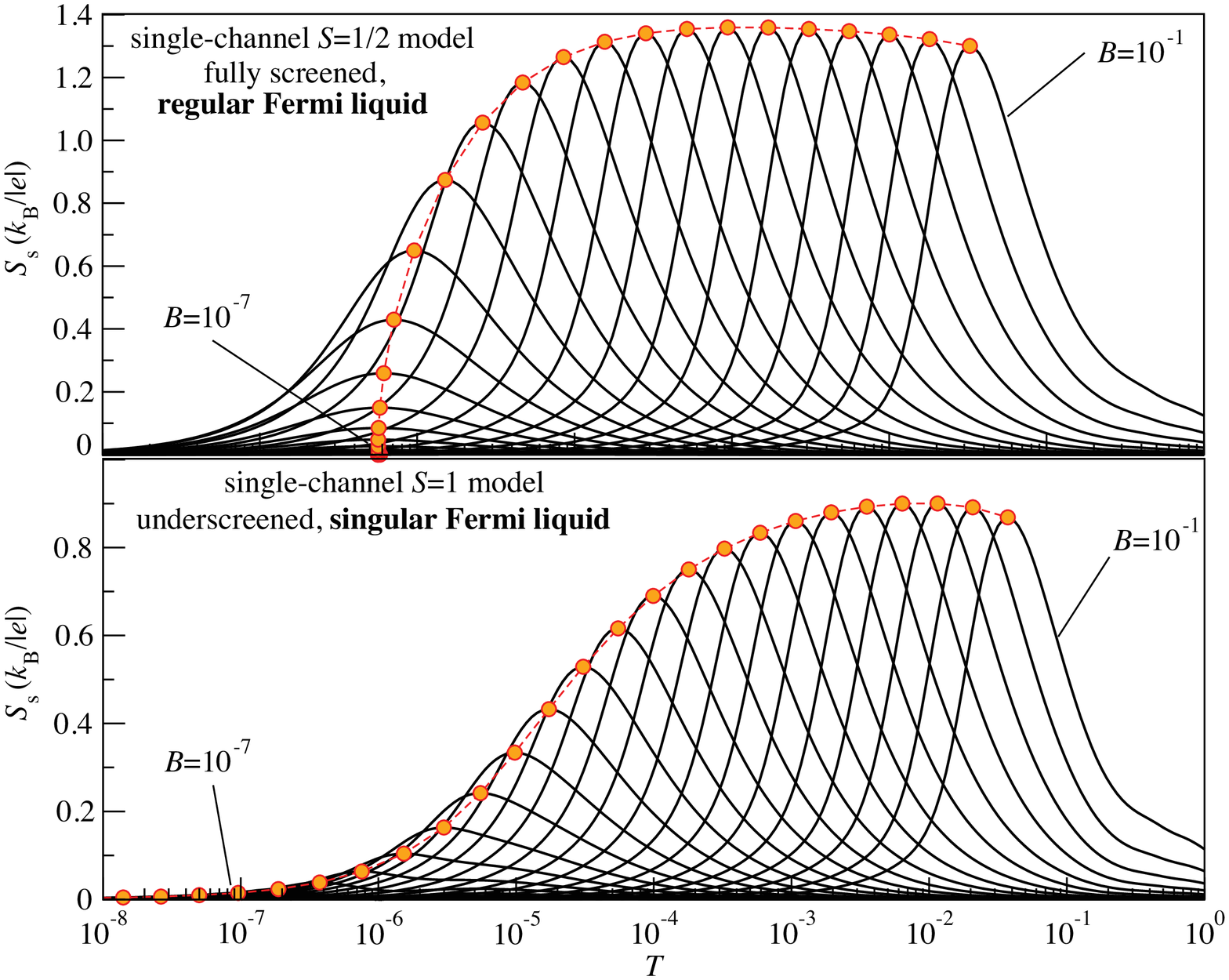}
\caption{Spin Seebeck coefficient for the single-channel Kondo model
with $S=1/2$ (top) and $S=1$ (bottom).
The magnetic field ranges from $B=10^{-7}$ to $B=10^{-1}$ in steps
of factor $10^{1/4}$.
}
\label{fig0b}
\end{figure}

\ack 

We acknowledge the support of the Slovenian Research Agency (ARRS)
under Program P1-0044.

\bibliography{stp2ch}

\end{document}